\documentclass[
 reprint,
 amsmath,amssymb,
 aps,
]{revtex4-2}

\usepackage{graphicx}
\usepackage{dcolumn}
\usepackage{bm}
\usepackage[caption=false]{subfig}
\begin{document}

\preprint{APS/123-QED}

\title{Controversy of quantum congestion in two-particle quantum walks}

\author{A.D. Lobanova}
\email{evstafeva.ad@phystech.edu}

\author{A.D. Lobanov}%
\email{lobanov.ad@phystech.edu}

\affiliation{%
 Moscow Institute of Physics and Technology, 9 Institutskiy per., Dolgoprudny, Moscow Region 141701, Russia.
}
\author{A.M. Pupasov-Maksimov}
\email{tretiykon@yandex.ru}
\affiliation{ Universidade Federal de Juiz de Fora
 Department of Mathematics}%


\date{\today}

\begin{abstract}
The article deals with one- and two-particle quantum walks on a graph with Braess-like topology and analyzes the issue of network congestion in the quantum world. Our approach to the study of congestion in quantum networks is based on the comparison of the evolution of bosonic and fermionic many-particle states. We consider a simple example of non-interacting particles, where one can expect the appearance of congestion in the fermion case due to the Pauli principle.  It is shown that dependence of the transport efficiency on the parameters of quantum graph is similar in the bosonic and fermionic cases.   
\end{abstract}

\maketitle

\section{\label{sec:level1}Introduction} 

\bibliographystyle{unsrt}

\title{Congested vs bare regimes of a multi-particle evolution on quantum graphs}
\author{Lobanova A.D., Pupasov-Maksimov A.M., Lobanov A.D.}
\date{July 2021}

\begin{figure*}
\includegraphics[scale=0.35]{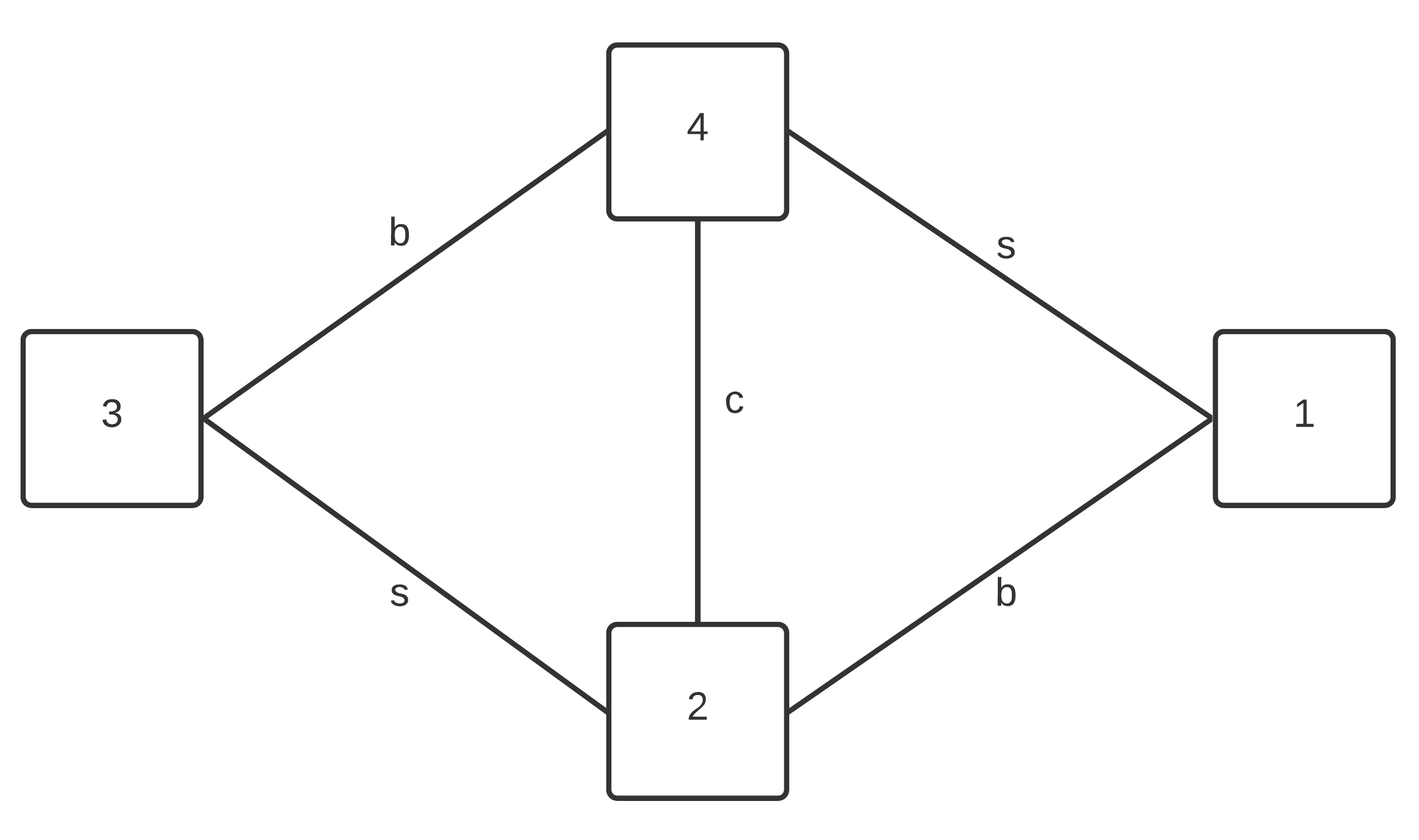}
\caption{A profile equivalent to the profile \cite{braess1968paradoxon} in terms of geometry.}\label{fig:mypic1}
\end{figure*}

\begin{figure*}
\includegraphics[scale=0.5]{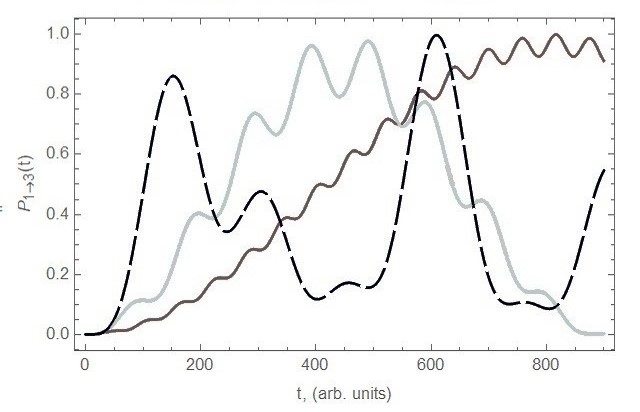}
\caption{Time dependence of the probability of transition from point 1 to point 3. The parameters of this system are shown in Table~\ref{tab:table1}. The black dashed line is $c = 0.01$, the gray line is $c = 0.05$, the black line is $c = 0.1$.}\label{fig:mypic2}
\end{figure*}

This paper is devoted to the study of the presence of Braess's paradox in the quantum world. This transport phenomenon has attracted a lot attention since the publication of the fundamental work of Braess \cite{braess1968paradoxon}. Braess's paradox is observed in transport networks of large cities, where road closures improve the situation on the roads \cite{zverovich2015braess}, or in electrical networks \cite{cohen1991paradoxical}; the paradox is also observed in computer science, in electric circuits \cite{nagurney2016observation} and in many other areas. 

In this paper, we study one- and two-particle quantum walks on a graph which is similar to the Braess transport network.
Scanning probe electronic imaging \cite{pala2012transport}, \cite{pala2012new}  indicates the existence of an analogue of Braess's paradox in branched-out
rectangular quantum rings \cite{martins2018scanning}. However, the issue of the  network congestion in the quantum theory is far from being clear. Such quantum phenomena as tunneling, spreading with time of quantum wave packets, state reduction, quantum statistics etc. make it difficult to use classical conceptions to interpret quantum transport efficiency. For instance, 
the question of tunneling time through a barrier has been a puzzle in foundational quantum mechanics for many decades \cite{spierings2020measuring,rivlin2021determination}. In \cite{lobanov2021twoparameter} we have shown that two mesoscopic samples with the same geometrical congestion (width of branches) and Braess-like topology demonstrate completely different behavior. It is an indication that the issue of network congestion which is crucial to the classical Braess paradox is non-trivial in quantum theory.

Braess's paradox appears as a result of a dynamical equilibrium in a multi-agent system. In the quantum multi-particle systems, the quantum statistics play an important role. If there is a limited number of available transverse modes in the mesoscopic sample  \cite{pala2012transport}, \cite{pala2012new}, it is natural to assume that the transport of fermions through such a conductor is congested due to the Pauli principle. 
Here we  study this assumption, isolating the possible effect of fermion statistics in a simple model. 
Our approach to the congestion of a quantum network is based on the comparison of the evolution of bosonic and fermionic many-particle (2-partical) states. Both the bosonic and fermionic versions of Braess's paradox are found in the proposed system of quantum dots (QDs) on a graph, which is quite surprising because any quantum graph should be un-congested to multi-particle bosonic quantum walks. 
The results of this paper can be used to analyze the speed of operation of computational elements (the rate of quantum evolution) based on QDs; the algorithm offers the opportunity to use the proposed characteristic of the time of passage of particles along a quantum graph for non-periodic evolution.

\section{Formulation of the transport problem}
Consider two-particle quantum walks on a graph. The system is a network of quantum dots (QDs) interconnected in a certain way - four double quantum dots (DQD) and two single quantum dots (see Fig. \ref{fig:mypic3}). Two non-interacting electrons move from the initial state to the final one. We study the dynamics of electrons in the absence of interaction between them. In experimental realizations, the Coulomb interaction of electrons at distant dots is suppressed by the large distance of interaction and by screening due to the presence of metallic gates between and nearby them \cite{melnikov2016quantum}.

Quantum dots can be formed from a two-dimensional electron gas by the gate field, and the population of electrons at these points can be controlled by the potentials at the gates. Each position on the graph can be occupied by no more than one electron. The position of the electron can be measured using quantum dot contact detectors, which are located near the quantum dots in such a way that the electron in a particular QD decreases the electric current in the detector, increasing the potential barrier. Therefore, a lower current detects an electron, and a higher current indicates an absence of an electron (empty QD), respectively. A system of semiconductor QDs is mathematically represented in the form of a graph, the vertices of which are QDs. The edges of the graph represent possible electron tunneling transitions. Electrons wander by tunneling through a barrier of controlled height between quantum dots. For simplicity, we will assume that the electron spins are always up (1/2), which can take place, for example, in a strong magnetic field \cite{melnikov2016quantum}.

Braess's paradox appears when an additional edge is provided to the transport network, however transportation time increases. Therefore we build two quantum graphs where the second one has an additional edge, as in \cite{pala2012new}, \cite{pala2012transport}. Then we compare the passage times between "initial" and "final" states for these systems. An increase of the passage time for the extended graph corresponds to an analogue of Braess's paradox. We will consider the transport flow of two fermions and two bosons. In the case of bosons, both quantum graphs are un-congested, and one can expect that an additional edge can only improve transport properties. 

\section{Non-interacting fermions on a quantum graph}

\subsection{Implementation of a transport one-particle system of quantum walks on a graph}

\begin{figure*}
\includegraphics[scale=0.3]{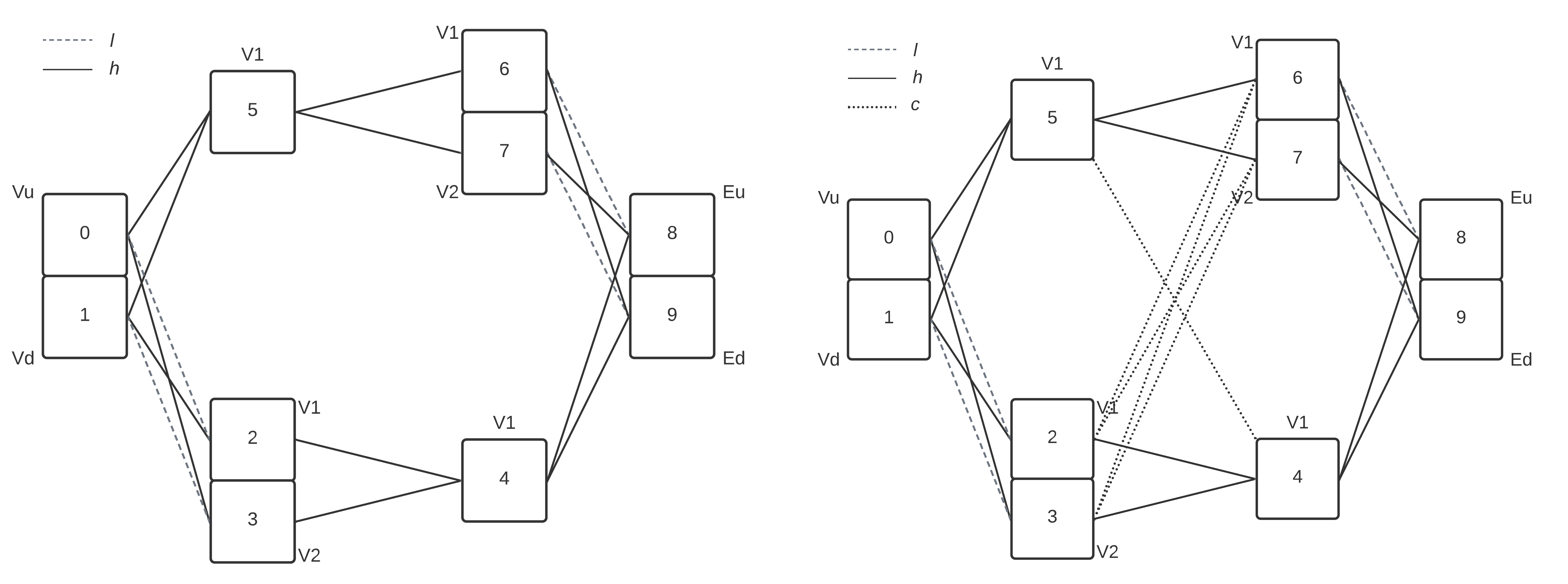}
\caption{Schematic representation of the systems. Coupling coefficients $ l, h $ and $ c $ are shown by different lines.}\label{fig:mypic3}
\end{figure*}

We start with a simple system of four single QDs (there is 1 bound state per dot), located at the vertices of the parallelogram (Fig.\ref{fig:mypic1}). Let the QDs be identical, and there are nonzero integrals of the overlap of states in pairs between all points.
We define the initial Hamiltonian of the system $H_{num}$ as follows:
\begin{equation*}
H_{num} =
\begin{pmatrix}
0 & b & 0 & s\\
b & 0 & s & c\\
0 & s & 0 & b\\
s & c & b & 0
\end{pmatrix}
\end{equation*}
Here, parameter $c$ regulates transition from QD 4 to QD 2, which is our additional edge in terms of Braess geometry. 
This Hamiltonian defines the evolution of a particle on the graph by the Sch\"{o}dinger equation. Eigenstates of $4\times 4$ Hamiltonian, which can be found analytically,  define time dependence of transition probabilities: 
$[1 \rightarrow 3]: P_{13}$, 
$[1 \rightarrow 2]: P_{12}$, 
$[1 \rightarrow 4]: P_{14}$, 
$[1 \rightarrow 1]: P_{11}$.
Below we consider a particular example. Numerical values of parameters are given in the Table~\ref{tab:table1}.
\begin{table}[b]
\caption{\label{tab:table1}%
Numerical values of coupling parameters and potential energy for a system of 4 QDs.
}
\begin{ruledtabular}
\begin{tabular}{cccc}
\textrm{b}&
\textrm{s}&
\textrm{c}&
\textrm{$V_{0}$}\\
\colrule
0.01 & 0.01 & 0.01, 0.05, 0.1 & 0\\
\end{tabular}
\end{ruledtabular}
\end{table}

Fig.~\ref{fig:mypic2} shows the time dependence of the transition probability $[1 (\text{initial}) \rightarrow 3 (\text{final})]$. One can observe that transition probability has a quasi-periodic character with a clearly detectable maximum probability of about 1. In this case, the time until transition probability reaches its maximum value close to 1 can be considered as a characteristic passage time. 

Addition of an edge $[4 \rightarrow 2]$ (parameter $c$) increases the oscillation period between $[1 \rightarrow 3]$ and thus slows down the evolution. However, it is clear that, in contrast to Braess's paradox, in this case no collective effects are observed, and the considered transport system is not congested. 
The change in the period of the oscillation is associated with a shift of eigenvalues under a perturbation due to an additional coupling. 
It is mostly a resonant effect originating from quantum scattering and interference \cite{sousa2013braess,barbosa2014universal}.

In the next section, we extend the model to introduce two-particle states. In this case, we can expect the emergence of collective effects. 

\subsection{Two-particle quantum walks on a graph}

We consider four double QDs and two single QDs which form a graph shown in Fig.~\ref{fig:mypic3}. 
The potential energies of the QD ${(V_u, V_d, V_1, V_2, E_u, E_d)}$ and the coupling coefficients between the QDs $(s, l, h, c)$ are schematically shown in Fig.~\ref{fig:mypic3}. 

The initial position is a double quantum dot with indices $[0,1]$, the finial position is a double QD with indices $[8,9]$. Initial and final QDs can be connected to external leads which provide initialization and measurement processes by charge-qubit operations as in \cite{gorman2005charge}. Note that we consider only evolution within the system of quantum dots (manipulation, in terms of \cite{gorman2005charge}). 
The initial state that we consider is a pure quantum state $|\psi(0)\rangle$, when both particles are in the initial double QD
\begin{eqnarray}
    |\psi(0)\rangle_f=\frac{1}{\sqrt{2}}(|0\rangle|1\rangle-|1\rangle|0\rangle)\,,
\\
    |\psi(0)\rangle_b=\frac{1}{\sqrt{2}}(|0\rangle|1\rangle+|1\rangle|0\rangle)\,.
\end{eqnarray}
In the absence of interactions, it is sufficient to determine one-particle evolution and then provide proper symmetrization thus obtaining $|\psi(t)\rangle$.  In the case of fermions, only one particle can occupy a single quantum dot due to the Pauli principle.
We also consider the same one-particle Hamiltonian to define the evolution of a pair of non-interacting bosons.

The characteristic of transport inefficiency is the probability $P_{\perp}(t)$ that both particles are restricted to the QDs from 0 to 7,
\begin{equation}\label{def:pperp}
    P_{\perp}(t)=\mathrm{Tr}\left(\hat{P}_{(0-7)}  |\psi(t)\rangle\langle \psi(t)|\right)=\langle \psi(t)|\hat{P}_{(0-7)}|\psi(t)\rangle\,,
\end{equation}
where the projector operator to the corresponding subspace
\begin{equation}\label{def:projperp}
    \hat{P}_{(0-7)}=\sum\limits_{i,j=0}^7 |i,j\rangle_{b,f}\langle i,j| \,,
\end{equation}
takes into account particle statistics.

\begin{figure*}
\centering
\subfloat[Subfigure 1 list of figures text][$l=0.1$, $h=0.2$, $s=0.25$, $V=0.5$]{
\includegraphics[width=0.4\textwidth]{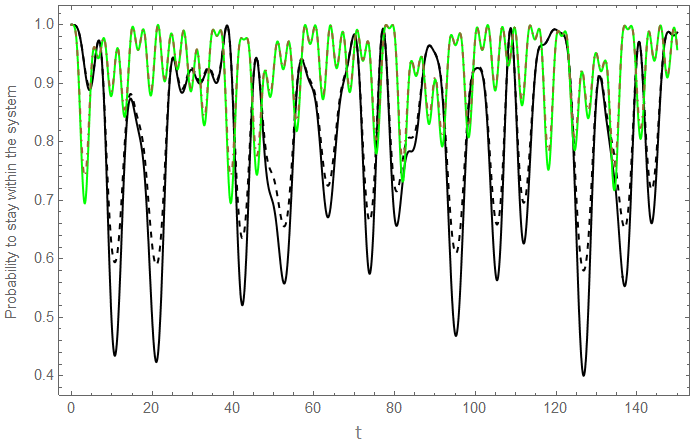}
\label{fig:mypic5}}
\qquad
\subfloat[Subfigure 2 list of figures text][$l=0.04$, $h=0.05$, $s=0.25$, $V=0$]{
\includegraphics[width=0.4\textwidth]{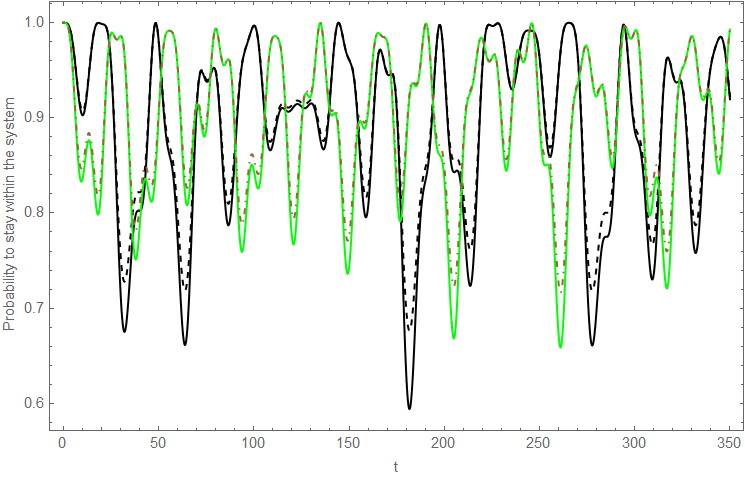}
\label{fig:mypic7}}
\caption{Time dependence of the probability $P_{\perp}(t)$ for the parameters of Tables~\ref{tab:table2} and \ref{tab:table3}, the coupling coefficient for the three paths for fermions and bosons is $c = 0.3$. The black line corresponds to two paths for fermions, the green line corresponds to three paths for fermions, the black dotted line corresponds to two paths for bosons and the brown dotted line corresponds to three paths for bosons.}
\label{fig:pperp}
\end{figure*}


Two numerical examples of $P_{\perp}(t)$ are given in Figs. \ref{fig:mypic5} and \ref{fig:mypic7}.
Parameters which correspond to Figs. \ref{fig:mypic5} and \ref{fig:mypic7} are shown in the Table~\ref{tab:table2} and Table~\ref{tab:table3}, respectively.
\begin{table}[b]
\caption{\label{tab:table2}%
First system parameters
}
\begin{ruledtabular}
\begin{tabular}{c|c|c|c|cccccc}
\textrm{l}&
\textrm{h}&
\textrm{s}&
\textrm{c}&
\textrm{$V_{u}|$}&
\textrm{$V_{d}|$}&
\textrm{$V_{1}|$}&
\textrm{$V_{2}|$}&
\textrm{$E_{u}|$}&
\textrm{$E_{d}$}\\
\colrule
0.1 & 0.2 & 0.25 & $\Delta$ & \multicolumn{6}{c}{0.5}\\
\end{tabular}
\end{ruledtabular}
\end{table}
\begin{table}[b]
\caption{\label{tab:table3}%
Second system parameters
}
\begin{ruledtabular}
\begin{tabular}{c|c|c|c|cccccc}
\textrm{l}&
\textrm{h}&
\textrm{s}&
\textrm{c}&
\textrm{$V_{u}|$}&
\textrm{$V_{d}|$}&
\textrm{$V_{1}|$}&
\textrm{$V_{2}|$}&
\textrm{$E_{u}|$}&
\textrm{$E_{d}$}\\
\colrule
0.04 & 0.05 & 0.25 & $\Delta$ & \multicolumn{6}{c}{0} \\
\end{tabular}
\end{ruledtabular}
\end{table}

We do not consider the following evolution, since it reduces to one particle case and possible time delays have a purely resonant character.
When $P_{\perp}(t)$ becomes small, at least one particle reaches final QD and can be measured there \cite{gorman2005charge}. 
The first non-intuitive result which is seen in these figures is that a pair of fermions will pass both quantum graphs faster $P_{\perp,f(t)}\leq P_{\perp,b(t)}$, despite the fact that there are single-particle channels where only one fermion can pass.
However this has a simple explanation, because the dimension of the two-particle subspace $V_{b}=\text{Span}\{|i,j\rangle_{b}, i,j=0,7\}$ of bosons is larger than the dimension of the two-particle subspace of fermions, $V_{f}=\text{Span}\{|i,j\rangle_{f}, i,j=0,7\}$, $\mathrm{dim}V_b=36>\mathrm{dim}V_f=28$. Note that boson and fermion Hamiltonians coincide in our model, and taking into account that the maximum number of mutually orthogonal states that the system can pass through per unit of time
is defined by the average energy E (above the ground state) \cite{margolus1998maximum}, the inequality $P_{\perp,f(t)}\leq P_{\perp,b(t)}$ follows.  

Probabilities $P_{\perp}(t)$ demonstrate chaotic behavior and it is difficult to compare which quantum graph has higher transport efficiency. An important issue which appears when we consider quantum transport is related to the probabilistic nature of quantum states.
In a simple graph considered earlier we saw almost periodic oscillations of probabilities. 
For instance, periodic dynamics in a ring of quantum dots were observed with up to 3 pairs of quantum dots in a ring \cite{melnikov2016quantum}.
In this case, the period of oscillations serves as a characteristic time of quantum evolution. In a more general case of quasi-periodic and chaotic dynamics of probabilities, there arises the problem of characteristic time.
We propose the following heuristic functional $\tau$ of the quantum evolution \begin{equation}\label{def:half-passage-time}
\lambda=\frac{1}{T}\int\limits_0^T\ln(P_{\perp}(t))dt\,,\qquad \tau=\frac{\ln 2}{\lambda}\,,
\end{equation}
which is called \textit{half-passage time}. If one considers an ensemble of identical quantum graphs with the same two-particle initial state, then the half-passage time corresponds to the time, when measurement in the final point will detect at least one particle in half of the graphs from the ensemble. 

\begin{figure*}
\centering
\subfloat[Subfigure 1 list of figures text][$l=0.1$, $h=0.2$, $s=0.01:0.01:0.3$, $V=0.5$]{
\includegraphics[width=0.4\textwidth]{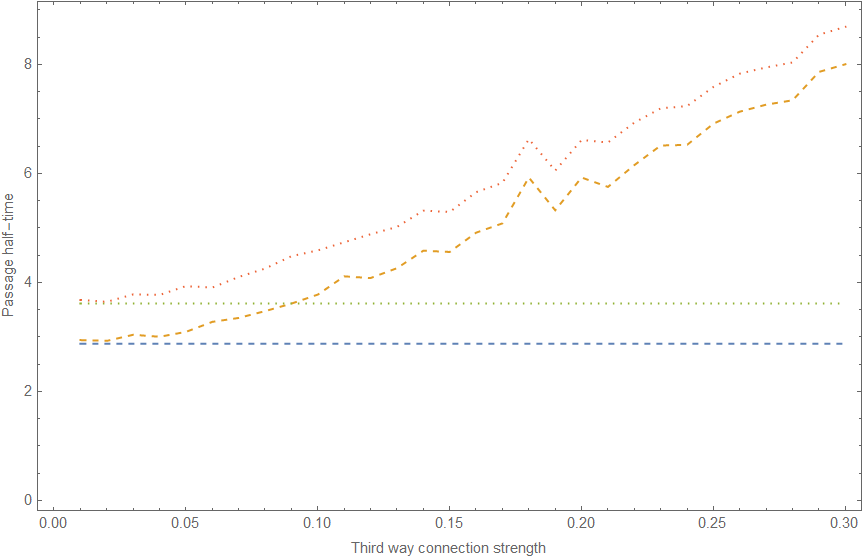}
\label{fig:mypic4}}
\qquad
\subfloat[Subfigure 2 list of figures text][$l=0.04$, $h=0.05$, $s=0.01:0.01:0.3$, $V=0$]{
\includegraphics[width=0.4\textwidth]{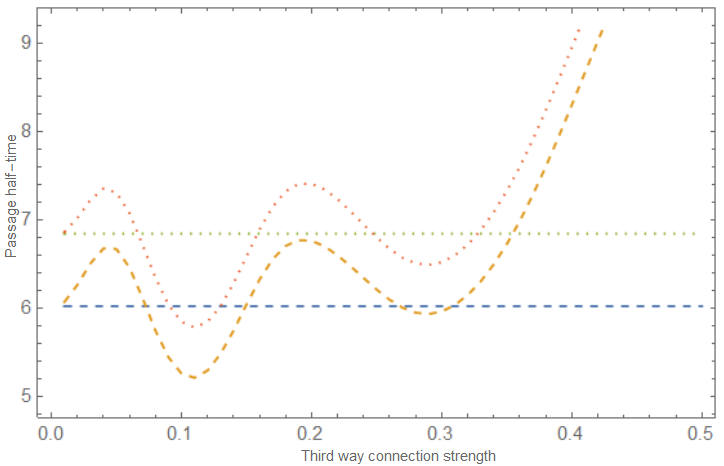}
\label{fig:mypic6}}
\caption{Dependence of the "half-passage" time $\tau$ on the coupling coefficient $c$ along the third path of the systems (Tables~\ref{tab:table2}, \ref{tab:table3} ). The blue horizontal dashed line - two paths for fermions, $c=0$. The orange dashed line - three paths for fermions. The green horizontal dotted line - two paths for bosons $c=0$. The red dotted line - three paths for bosons.}
\label{fig:half-passage}
\end{figure*}

In 
Figs.~\ref{fig:mypic4}, \ref{fig:mypic6} we plot 
the "half-passage" time $\tau$ as a function of the coupling coefficient $c$, where the initial state is $|0,1\rangle$, in the initial double quantum dot. In 
Fig.~\ref{fig:mypic4} the half-passage time increases in both boson and fermion cases which is similar to the increasing period of oscillations in Fig.~\ref{fig:mypic2}. Since bosons do not have congested channels, we can conclude that Braess-like behavior is not related to the congestion of the network. It is mostly a single-particle resonant effect. 
In the case of weakly connected quantum graph, Fig. \ref{fig:mypic6}, it is seen that there occur both Braess-like and "normal" behavior of transport efficiency measured by $\tau(c)$. Again, large $c$ looks like Braess's paradox, which has a resonant character (without involving multi-particle effects). Note that half-passage time is longer in the boson case, $\tau_{b}>\tau_{f}$, again due to the larger dimension of the corresponding two-particle Hilbert space. 

\section{Conclusion}
This paper considered a mathematical model of one-particle and two-particle quantum walks on a graph. Numerical modeling has shown that, in the simplest Braess topology, the inclusion of an additional coupling between QDs leads to an increase in the oscillation period (evolution deceleration). Apparently, the reason for the quantum analogue of Braess's paradox is a change in the spectrum of the system, in which bound states of an electron inside the system slow down its evolution (i.e., it has a resonance capture character, see also \cite{sousa2013braess,barbosa2014universal}).

To study the effect of Pauli exclusion principle on network congestion, we considered a two-particle evolution. 
In the case of quasi-periodic and chaotic time dependence of transition probabilities, half-passage time \eqref{def:half-passage-time} can be used to measure transport efficiency. It corresponds to the time, when measurement at the final point will detect at least one particle in half of the graphs from the ensemble. Our numerical experiments show that pairs of bosons or fermions demonstrate similar transport behavior in quantum graphs. Therefore, we have come to the conclusion that the Pauli exclusion principle for a small-sized graph does not lead to the congestion of the system. Nevertheless, a Braess-like behavior of transport efficiency reduction is observed both for bosons and fermions for some parameters. 


\bibliography{spin}

\end{document}